\documentclass[twocolumn,aps,prl,showpacs,amsmath,amssymb,superscriptaddress]{revtex4}

\usepackage{graphicx}
\usepackage{amsmath,amssymb}
\usepackage{textcomp}

\newcommand{\ket}[1]{\ensuremath{\left|  #1 \right\rangle}}

\newcommand{\aver}[1]{\ensuremath{\langle  #1 \rangle}}
\newcommand{\hh}[0]{\ensuremath{\ket{\textrm{h}}}}
\newcommand{\vv}[0]{\ensuremath{\ket{\textrm{v}}}}
\newcommand{\CSS}[0]{\ensuremath{\phi_{css}}}

\begin{document}

\title{Generating Entangled Spin States for Quantum Metrology by Single-Photon Detection}

\author{Robert McConnell}
\affiliation{
Department of Physics, MIT-Harvard Center for Ultracold Atoms,
and Research Laboratory of Electronics, Massachusetts Institute of Technology,
Cambridge, Massachusetts 02139, USA}

\author{Hao Zhang}
\affiliation{
Department of Physics, MIT-Harvard Center for Ultracold Atoms,
and Research Laboratory of Electronics, Massachusetts Institute of Technology,
Cambridge, Massachusetts 02139, USA}

\author{Senka \'{C}uk}
\affiliation{
Department of Physics, MIT-Harvard Center for Ultracold Atoms,
and Research Laboratory of Electronics, Massachusetts Institute of Technology,
Cambridge, Massachusetts 02139, USA}
\affiliation{
Institute of Physics, University of Belgrade, Pregrevica 118, 11080 Belgrade, Serbia}

\author{Jiazhong Hu}
\affiliation{
Department of Physics, MIT-Harvard Center for Ultracold Atoms,
and Research Laboratory of Electronics, Massachusetts Institute of Technology,
Cambridge, Massachusetts 02139, USA}

\author{Monika H. Schleier-Smith}
\affiliation{
Max-Planck-Institut f{\"u}r Quantenoptik, Ludwig-Maximilians-Universit{\"a}t, Schellingstr. 4, 80799 M{\"u}nchen}

\author{Vladan Vuleti\'{c}}
\affiliation{
Department of Physics, MIT-Harvard Center for Ultracold Atoms,
and Research Laboratory of Electronics, Massachusetts Institute of Technology,
Cambridge, Massachusetts 02139, USA}


\date{\today}

\begin{abstract}
We propose and analyze a probabilistic but heralded scheme to generate pure, entangled, non-Gaussian states of collective spin in large atomic ensembles by means of single-photon detection. One photon announces the preparation of a Dicke state, while two or more photons announce Schr{\"o}dinger cat states. The method produces pure states even for finite photon detection efficiency and weak atom-photon coupling. The entanglement generation can be made quasi-deterministic by means of repeated trial and feedback, enabling metrology beyond the standard quantum limit.
\end{abstract}

\maketitle

State-of-the-art atomic clocks and other atom interferometers are limited by quantum projection noise. For measurements on a system of $N$ uncorrelated atoms in a coherent spin state (CSS), this projection noise sets a limit scaling as $1/\sqrt{N}$, referred to as the standard quantum limit (SQL). Entangled states can overcome this limit, potentially reaching the Heisenberg limit, where uncertainty scales as $1/N$. Thus far, the potential for metrological gain has been demonstrated in atomic ensembles using squeezed spin states \cite{Kitagawa1993, Appel09,Takano10,Schleier-Smith10,Leroux2010, Chen2011,Gross2010, Riedel2010, Hamley12}, which have enabled atomic clock operation surpassing the SQL \cite{Louchet2010, LerouxSqueezedClock2010}. In these experiments, the entanglement has been produced either by spin-dependent atom collisions \cite{Gross2010, Riedel2010,Hamley12}, or by coupling an optical probe to the atomic ensemble \cite{Appel09,Takano10,Schleier-Smith10, Leroux2010, Chen2011}. The Greenberger-Horne-Zeilinger states \cite{GHZ1990} have also been shown to allow metrological gain \cite{BollingerNOON1996}, and have been produced \cite{Leibfried2004,Roos2004} for collections of up to 14 ions \cite{Monz11} via Coulomb interactions.

In this Rapid Communication, we describe a method to generate pure entangled states of collective spin in large atomic ensembles for measurements beyond the SQL. Photons transmitted through the ensemble experience a weak random Faraday rotation associated with the quantum noise of the atomic spin. A photon emerging with polarization orthogonal to its input polarization heralds the creation of a non-Gaussian entangled state of collective atomic spin, and two or more orthogonally polarized photons herald increasingly more entangled ``squeezed Schr\"odinger cat'' states \cite{Brask2010}. This method generates states of high purity even for weak atom-photon coupling and finite photon detection efficiency, which simply reduce the probability of entangled-state preparation. The heralded entanglement scheme can be made quasi-deterministic through repeated trial and feedback, enabling atom interferometry beyond the SQL.

\begin{figure}
\centering
\includegraphics[width=.45 \textwidth]{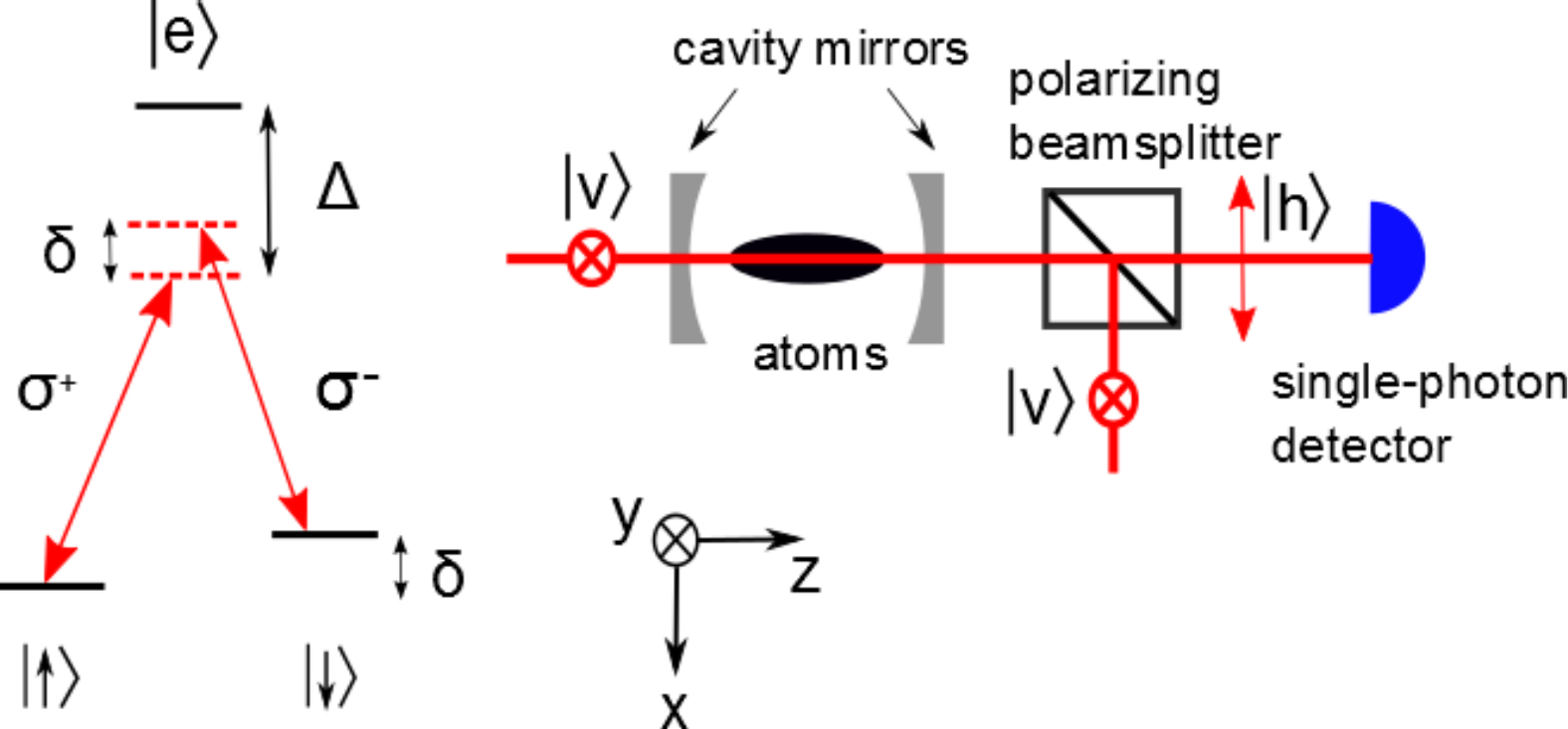}
\caption{Scheme for the heralded generation of nonclassical states. (a) Atoms with two spin states $\ket{\uparrow}$ and $\ket{\downarrow}$ are coupled to an electronic excited state $\ket{e}$ via two degenerate circularly polarized modes. (b) Incident vertically polarized photons experience weak Faraday rotation as they traverse the ensemble. The detection of a horizontally polarized transmitted photon heralds the generation of a non-Gaussian entangled state of collective atomic spin. A cavity enhances the Faraday rotation and the state preparation probability.}
\label{fig:Fig1}
\end{figure}

Our approach has similarities to weak-measurement schemes \cite{Aharonov88} that use postselection to enable detection of small signals in the presence of technical noise \cite{Hosten08, Dixon09, Simon11, Brunner11}. It is closely related to heralded schemes for the creation of Dicke states in atomic ensembles for quantum communication \cite{DLCZ2001} which have been experimentally implemented for photon-pair and single-photon generation \cite{Kuzmich03,Matsukevich06,Simon07}. In our method, quantum noise of the atomic state produces a weak Faraday rotation of the polarization of a photon, whereby the phase of the atomic CSS becomes entangled with the photon polarization. The detection of a single photon of select polarization then prepares the atomic ensemble in a non-Gaussian entangled state that results from destructive interference between two weakly separated coherent states. This method can be implemented either in free space or in an optical cavity; the latter increases the polarization rotation and hence the entanglement rate. When the state preparation is fast compared to the atomic coherence time, as is the case in atomic clocks and many interferometers, then the present method can be made quasi-deterministic by repeated trial and feedback, enabling interferometry beyond the SQL. The metrological gain is 3dB for just one detected photon, and improves with additional detected photons. We note that related methods to generate squeezed Gaussian states by measuring the Faraday rotation of a light pulse containing a large number of photons have been proposed \cite{Kuzmich1998,Takahashi1999,Vanderbruggen11} and implemented \cite{Takano10}. The scheme proposed in \cite{Christensen2012} for Dicke states is similar to our scheme and should allow the same metrological gain.

Consider an ensemble of $N$ three-level atoms. Two ground states $\ket{\uparrow},  \ket{\downarrow}$, e.g. magnetic sublevels, correspond to a pseudo-spin $s_i=\frac{1}{2}$. The collective state of the ensemble can be described by a total spin $\bf{S}=\sum \bf{s}_i$ that is the sum of individual spins $\bf{s}_i$. Two degenerate, oppositely circularly polarized modes of an optical cavity couple $\ket{\uparrow}$ and $ \ket{\downarrow}$ to an excited state $\ket{e}$ (Fig.~\ref{fig:Fig1}). A magnetic field applied along the quantization axis $\hat{z}$ lifts the degeneracy between the ground states by an amount $\hbar \delta$ such that the two-photon Raman coupling between $\ket{\uparrow}$ and $\ket{\downarrow}$ is negligible. We assume that the two transitions have equal coupling strength and that all atoms are equally coupled to the light, with single-photon Rabi frequency $2g$. When the light-atom detuning $\Delta$ is much larger than the excited-state width $\Gamma$, we can adiabatically eliminate the excited state $\ket{e}$. Ignoring photon emission into free space for now, the interaction Hamiltonian for the atom-photon system is written as \cite{Kuzmich1998}
\begin{equation}
\frac{H}{\hbar}= \left( \frac{2 g^2}{\Delta} \right)  J_z S_z.
\label{eqn:Hamil}
\end{equation}
$\bf{J}$ is the Stokes vector of light and obeys the commutation relation of angular momenta $[J_i, J_j]=i \epsilon_{ijk}J_k$. In particular, $J_z= \frac{1}{2}(a^\dagger_+ a_+ -a^\dagger_- a_-)$ where $a_\pm$ are the annihilation operators of $\sigma^{\pm}$ light.
The atoms are prepared initially in the CSS $\ket{x}$ along $\hat{x}$, satisfying $S_x |x \rangle = S |x \rangle$ where $S=N/2$. Consider a vertically polarized incident photon described by the state $\vv = (\ket{\sigma^+} +\ket{\sigma^-})/\sqrt{2}$. While the photon is inside the cavity the atom-photon system evolves as $e^{-iHt} \ket{x} \vv$, which after the photon has been transmitted through the cavity results in the state \cite{Leroux12}
\begin{eqnarray}
\nonumber  \ket{\psi_t} &=& \frac{1}{\sqrt{2}} \sum_{m=-S}^{S} c_m \ket{m} (e^{-im\phi} \ket{\sigma^+} +e^{im\phi} \ket{\sigma^-}) \\
  &=& \frac{1}{\sqrt{2}} \left( \ket{\sigma^+} \ket{\phi} + \ket{\sigma^-} \ket{-\phi} \right).
  \label{eqn:evolve}
\end{eqnarray}
Here $\phi= \eta \Gamma / 2 \Delta$ is an accumulated phase, expressed in terms of the cavity linewidth $\kappa$ and the single-atom cooperativity $\eta = 4 g^2/\kappa \Gamma$ \cite{Leroux12,Tanji-Suzuki11a}, and the atomic state is written in terms of $S_z$ eigenstates $\ket{m}$ and binomial coefficients $c_m = 2^{-S} [(2S)!/((S+m)!(S-m)!)]^{1/2}$. Here $\ket{\pm \phi}$ designates the CSS in the equatorial plane rotated by an angle $\pm \phi$ about $\hat{z}$ away from $\hat{x}$. In the following we restrict the analysis to weak atom-cavity coupling, $\eta \ll 1$, and the dispersive limit of low photon absorption, requiring \cite{Tanji-Suzuki11a} $2S \eta (\Gamma/2 \Delta)^2 \ll 1$. This implies that the angle $\phi$ is much smaller than the CSS angular width $\CSS=1/\sqrt{2S}$.

\begin{figure}[h!tbp]
\centering
\includegraphics[width=.45 \textwidth]{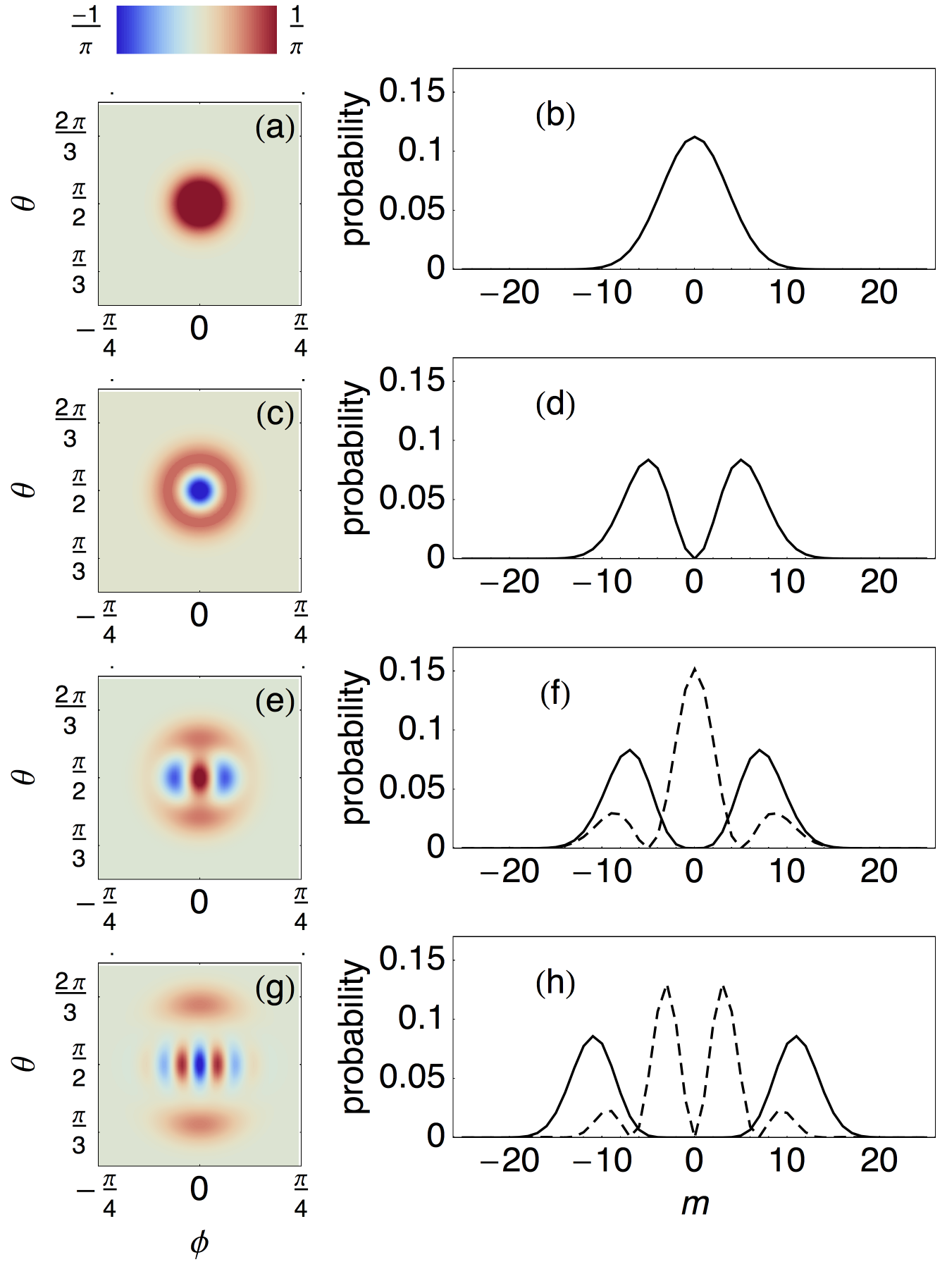}
\caption{Normalized Wigner quasiprobability distribution $W(\theta , \phi)/\hbar \sqrt{S(S+1)}$ (left) and probability distributions of angular-momentum eigenvalues (right, solid line for $S_z$, dashed line for $S_y$) calculated for $N = 50$ atoms for ((a), (b)) the input CSS, ((c), (d)) $n=1$ detected $\hh$ photon, ((e), (f)) $n=2$ detected $\hh$ photons, ((g), (h)) $n=5$ detected $\hh$ photons. Wigner functions (e), (g) indicate the production of a Schr{\"o}dinger cat state for $n \geq 2$ detected $\hh$ photons. The distributions of $S_y$ eigenvalues in ((d), (f), (h)) consist of several peaks, narrower than the CSS width, enabling measurements surpassing the SQL.}
\label{fig:stateplots}
\end{figure}

The first line of Eq.~\ref{eqn:evolve} is readily interpreted as the phase $\pm m \phi$ being imprinted onto the $\sigma^{\pm}$ polarizations of the light field due to the refractive index of the atoms in the states $\ket{\uparrow}, \ket{\downarrow}$ with population difference $2m$. A value $m \neq 0$, i.e. a deviation of $S_z$ from its mean value $\aver{S_z}=0$ due to quantum noise in the atomic state, thus results in a polarization rotation of the photon.  The detection of a horizontally polarized photon $\hh$ requires $S_z \neq 0$, and biases the system towards states with larger $|S_z|$, creating a collective spin state whose quasiprobability distribution on the Bloch sphere shows a hole in the center (see Fig. \ref{fig:stateplots}).

From a complementary viewpoint, $\sigma^+$ and $\sigma^-$ photons shift the phase of the atomic CSS in opposite directions by an amount $\pm \phi$. Even though $\phi \ll \CSS$, the detection of a horizontally polarized photon $\hh=(\ket{\sigma^+} - \ket{\sigma^-})/\sqrt{2}$ corresponds (according to the second line in Eq.~\ref{eqn:evolve}) to the destructive interference $\ket{\phi}-\ket{-\phi}$ of two weakly separated coherent spin states, which generates the hole in the center of the state.

The atomic state after detection of one photon in $\hh$ can be expressed as $\ket{\psi_1} = \sum_{m}c_m\sin(m \phi)\ket{m}$, where the unimportant normalization factor has been omitted, and for $\phi \ll \CSS$ approximated as $\ket{\psi_1} = \sum_{m}m c_m \ket{m}$. $\ket{\psi_1}$ is the first Dicke state along $\hat{x}$, which satisfies $S_x \ket{\psi_1} = (S-1)\ket{\psi_1}$. Now consider an input Fock state of $n_0$ photons, with $n$ photons exiting the system in $\hh$ and $n_0-n$ photons exiting in the original polarization $\vv$. The atomic state is then given by
\begin{equation}
|\psi_n \rangle=\displaystyle\sum_{m} c_m \sin^{n}(m \phi)\cos^{n_0-n}(m \phi)\ket{m}.
\label{eqn:poststate2}
\end{equation}
In the dispersive limit and for small $\eta$, $ \cos (m\phi) \approx 1$ for $m \lesssim \sqrt{S/2}$, and the state $|\psi_n \rangle$ for $n \geq 2$ corresponds to a ``squeezed cat'' state \cite{Brask2010}, a superposition of two Gaussian states squeezed by a factor of 2 and separated on the Bloch sphere by an angle $\Delta \phi = 2 \sqrt{n / S} = \sqrt{8n} \CSS$  (see Fig. \ref{fig:stateplots}). Remarkably, as both the separation angle $\Delta \phi$ and the CSS angular width $\CSS$ scale as $1/\sqrt{S}$, this allows the production of states separated by an angle greater than the CSS width for just a few detected photons, regardless of the atom number used. We emphasize that for $\phi \ll \phi_{css}$ the states $\ket{\psi_n}$ are independent of $\phi$, which affects only the likelihood of producing the state.

The entangled states $\ket{\psi_n}$ display peaks in the angular-momentum distributions along both $S_z$ and $S_y$ that are narrower than the CSS width. In particular, expressed in terms of $S_y$ eigenstates $\ket{m}_y$, the state $\ket{\psi_n}$ is to lowest order independent of $n_0$ and is well approximated by
\begin{equation}
\ket{\psi_n} = \left\{
        \begin{array}{ll}
            \displaystyle\sum_m A_{n,S} e^\frac{-m^2}{4 S} \sin (m \sqrt{n/S}) \ket{m}_y & \quad n \, \, \textrm{odd} \\
            \displaystyle\sum_m A_{n,S} e^\frac{-m^2}{4 S} \cos (m \sqrt{n / S})  \ket{m}_y & \quad n \, \, \textrm{even}
        \end{array}
    \right.
    \label{eqn:Sy}
\end{equation}
where $A_{n,S} = (\pi S/2)^{-1/4} (1-e^{-2 n})^{-1/2}$ is a normalization constant. Figure 2 shows the $S_z$ and $S_y$ probability distributions of the state $\ket{\psi_n}$ and the corresponding normalized Wigner function, $W(\theta , \phi)/\hbar \sqrt{S(S+1)}$, for the spin state \cite{Dowling1994}. Figure \ref{fig:stateplots}(a)-(b) shows the input CSS, while Figure \ref{fig:stateplots}(c)-(d) shows the state produced by the conditional detection of one photon in $\hh$. Higher-order states produced by the conditional detection of more than one $\hh$ photon are shown in Figure \ref{fig:stateplots}(e)-(h).

The narrower features along $S_y$ enable improved phase readout in a Ramsey measurement compared to the CSS: After initial state preparation, a period $\tau$ of free evolution is followed by rotation about the average direction of the spin vector, which can be chosen as $\hat{x}$, thereby mapping the multipeaked $S_y$ distribution onto the $S_z$ axis. The value of $\aver{S_z}$, and hence the accumulated interferometer phase, is found by fitting the measured distribution of $S_z$ values to the \emph{a priori} distribution given by Eq. \ref{eqn:Sy}. (We assume that decoherence in the interferometer leads to phase fluctuations much less than the width of the peaks.) To see that this procedure gives lower quantum noise in the measurement of $S_z$, consider $M$ measurement points, of which a known fraction $f_i$  falls under a particular peak $i$, $\mu_i$ is the mean value of $S_z$ associated with that peak and $\sigma_0$ is the width of each peak. The weighted average, given by $S_z = \sum_i {\mu_i f_i}$, has variance given by $(\Delta S_z)^2 = \sigma_0^2 / M$, the same as for $M$ measurements conducted on a single peak of width $\sigma_0$. Thus, a probability distribution composed of multiple narrow peaks allows the same reduction in measurement uncertainty as one containing a single peak of equal (reduced) width. This allows the entangled states $\ket{\psi_n}$ to produce substantial metrological gain. In particular, the first excited Dicke state, produced by a single detected photon, results in measurement variance 3.4 dB below the SQL. This metrological gain is confirmed by calculations of the classical Fisher information in the $S_z$ distributions \cite{Giovannetti2011}, which show enhancement beyond the SQL in agreement with the values obtained by the measurement protocol we have described.
\begin{figure}[htbp]
\centering
\includegraphics[width=.45 \textwidth]{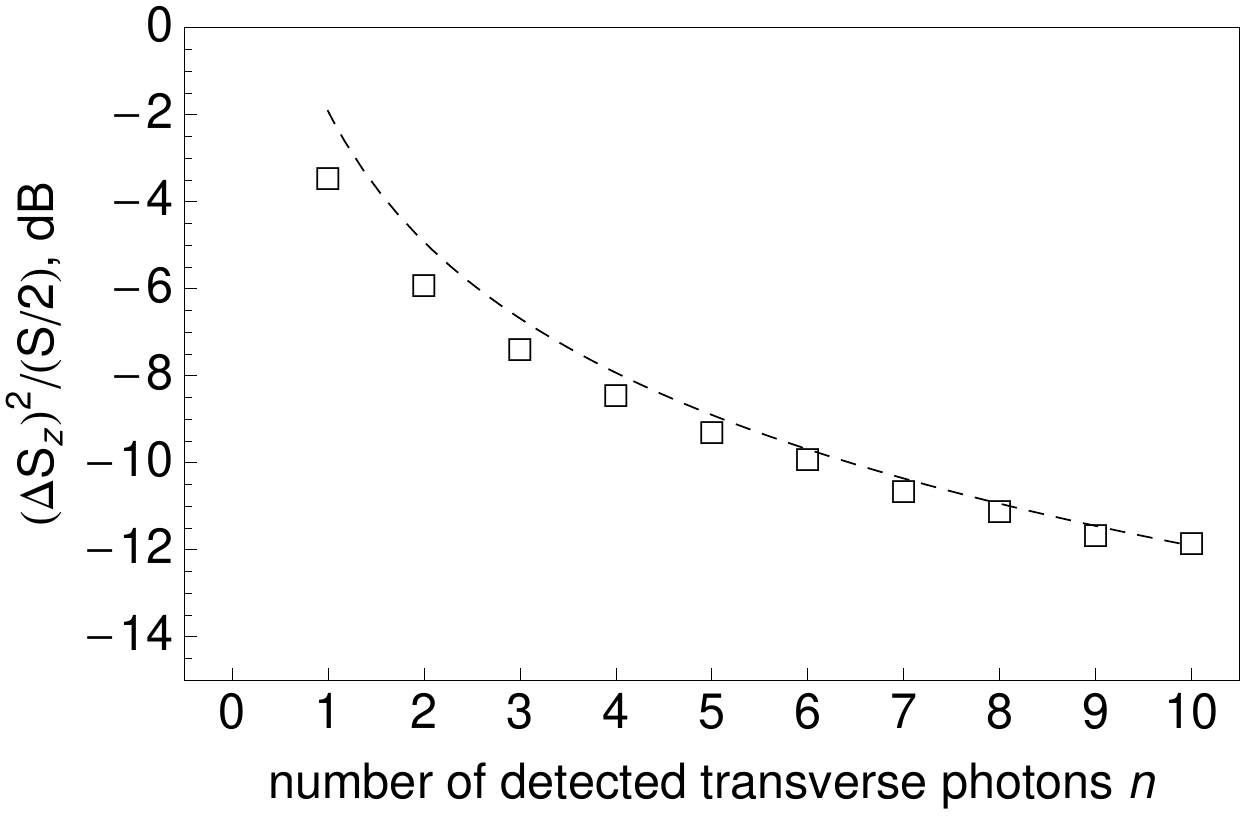}
\caption{(Black squares) Measurement variance of the states $\ket{\psi_n}$, normalized to the CSS variance, $(\Delta S_z)^2 / (S/2)$, in dB, as a function of number of detected $\hh$ photons $n$, indicating substantial metrological gain for $n$ of a few. (Dashed curve) For larger $n$ the normalized variance asymptotically approaches the value $(\Delta S_z)^2 / (S/2) = 0.64 /n$ (see text). Due to the finite atom number, $N=100$, used for the calculations, points in the figure show slight deviations from the asymptotic behavior expected for large $N$.}
\label{fig:metrogain}
\end{figure}

Figure \ref{fig:metrogain} shows the measurement variance (normalized to the CSS variance) as a function of detected photon number $n$, assuming large atom number $N$. The normalized variance asymptotically approaches the value $0.64 /n$ (dashed curve in Fig. \ref{fig:metrogain}). This represents the squared ratio of the width of one peak of the function $\cos^2(m \sqrt{n/S})$, assuming $n$ large, to the CSS width $\sqrt{S/2}$. While the probability to produce high-$n$ states decreases exponentially, Fig. \ref{fig:metrogain} indicates that substantial metrological gain is obtained even for $n$ of a few. (Similar calculations for higher-order excited Dicke states \cite{Christensen2012} indicate that the $n$-th Dicke state results in the same metrological gain as the $n$-photon state produced by our scheme.)

To calculate the heralded generation rate of these entangled states, we note that the probability of converting one incident $\vv$ photon into an $\hh$ photon and detecting it is easily calculated from the mean square of the polarization rotation angle $\aver{\beta^2} = S \phi^2 /2$, and is given by $p= q S \phi^2 /2 \ll 1$, where $q\leq 1$ is the photon detection efficiency. The probability of the incident photon being scattered into free space by the atomic ensemble is $p_{\mathrm{sc}}= 2S \eta(\Gamma/2\Delta)^2 = 2 S \phi^2 / \eta$ \cite{Tanji-Suzuki11a}. Therefore the success probability is simply related to the free-space scattering probability via $p = q \eta p_{\mathrm{sc}}/4$. A cavity increases the single-atom resonant optical depth $2 \eta$ \cite{Tanji-Suzuki11a} and hence greatly improves the generation efficiency for a given $p_{\mathrm{sc}}$. 

The input photons will typically be in a coherent state with mean photon number $n_0$. Photons exiting the system in the original polarization $\vv$, whether detected or not, have minimal impact on the atomic state: they multiply the coefficients $c_m$ of the state by $ \cos (m \phi) \approx 1$ as shown in (\ref{eqn:poststate2}). The series of cosine factors does not significantly degrade the atomic state until $N n_0 \phi^2$ approaches unity. Since the probability $p_1$ to detect one outgoing $\hh$ photon out of $n_0$ incident $\vv$ photons is given by $p_1 = q N n_0 \phi^2/4$, this results in the requirement $p_1 \ll q$.

The produced state can also be degraded by undetected $\hh$ photons. When a coherent state is used for the input, the probability to detect exactly $n$ photons in $\hh$ is given by the weighted sum over values of $S_z$ of $P(\bar{n}(S_z), n)$, where $P(\bar{n}(S_z), n)$ is the chance to find exactly $n$ output photons given a Poisson distribution with mean value $\bar{n}(S_z)=q n_0 S_z^2 \phi^2$. For $p_1 \ll 1$, the overall probability for exactly $n$ photons to exit the system in $\hh$ is $(p_1/2)^n (2n)!/ (n!)^2 q^n$, and the probability to detect them all is $p_n=(p_1/2)^n (2n)!/ (n!)^2 $. The probability that $n+1$ photons exit in $\hh$ of which $n$ are detected is then $p_{n+1} q^n (1-q) (n+1)/q^{n+1} = p_1 p_n (2 n + 1) (1-q)/q$. Such ``false positive'' states, corresponding to an additional undetected $\hh$ photon, produce an atomic state different from the heralded state, substantially reducing the signal-to-noise ratio. Under the requirement $p_1 \ll q$, the probability for such ``false-positive'' states is smaller than that of the heralded state by a factor $p_1 (2 n +1 )(1-q)/q$.

Note that, in order to maintain the coherence of the atomic spin state, the photon number scattered into free space must remain substantially smaller than the atom number, $n_0 p_{sc} \ll N$. For optically dense ensembles in free space, $2 N \eta >1$, this condition is automatically met by $p_1 \ll 1$, and the method proposed here can also be directly applied to dense ensembles in free space.

Given a coherent input state with mean photon number $n_0$, the probability per trial to detect $n$ photons is $(p_1/2)^n (2n)!/ (n!)^2$, which requires on average only a small number $n_{sc} = 4 p_1 / q \eta$ of photons to be scattered into free space. While the success probability decreases exponentially with $n$, even states corresponding to $n$ of a few display significant non-classicality. For instance, with a realistic $q=0.5$ detection efficiency and choosing $n p_1=0.2q$, the creation of entangled states corresponding to $n=1,2,3$ requires on average 10, 300, and $1 \times 10^4$ trials, respectively, for any number of atoms. The corresponding improvements over the SQL are 3.4 dB, 6.0 dB and 7.4 dB, respectively.

Although the states are generated only probabilistically, due to the heralding, preparation attempts may be repeated until success. For Ramsey measurements, the free precession time $\tau$ is typically much longer than the state preparation time. Under these conditions, even if entangled state preparation requires many attempts, the total preparation time can remain small compared to $\tau$ and there is no significant reduction of measurement duty cycle. As one example, state preparation in the Sr optical lattice clock is limited by the 20 $\mu$s decay time of the optical pumping transition, suggesting that state preparation, consisting of optical pumping to $\ket{\downarrow}$, preparation of a coherent state $\ket{x}$ with a $\pi/2$ pulse, and illumination by the probe pulse, could be performed in as little as 200 $\mu$s, while the available measurement time for that system is $\sim$ 1~s and is currently laser-limited \cite{Katori2011}. Thus, up to $\sim 10^4$ state preparation attempts may be made without compromising available measurement time; a transition with faster excited-state decay time may allow even more attempts.

In conclusion, we have proposed a technique where a single photon can create an entangled spin state of a very large ensemble of atoms. This can be achieved even in the limit of weak coupling between a photon and an atom and finite photon detection efficiency. The use of these states for interferometry below the SQL requires state readout capabilities well below the CSS width, as have been recently demonstrated \cite{Zhang2012}.

This work was supported by the NSF, DARPA (QUASAR) and MURI through ARO, and the NSF Center for Ultracold Atoms. S.\'{C}. acknowledges support from the Ministry of Education, Science and Technological Development of the Republic of Serbia, through Grant No. III45016.

\end{document}